\documentclass{amsart}

\theoremstyle{definition}

\theoremstyle{remark}

\numberwithin{equation}{section}

\begin{document}

\title[A novel weighting scheme for random $k$-SAT]{A novel weighting scheme for random $k$-SAT}

\author{Zongsheng Gao}

\address{Zongsheng Gao: LMIB and School of Mathematics and Systems Science, Beihang University, Beijing, 100191, P.R. China}

\email{zshgao@buaa.edu.cn}

\author{Jun Liu}
\address{Jun Liu: LMIB and School of Mathematics and Systems Science, Beihang University, Beijing, 100191, P.R. China}
\email{junliu@smss.buaa.edu.cn}
\thanks{This research was supported by National Natural Science Fund of China (Grant No.11171013, 60973033).}

\author{Ke Xu}
\address{Ke Xu: State Key Laboratory of Software Development Environment, Department of Computer Science, Beihang University, Beijing, 100191, P.R. China}
\email{kexu@nlsde.buaa.edu.cn}


\date{}

\dedicatory{}


\begin{abstract}
Consider a random $k$-CNF formula $F_{k}(n, rn)$ with $n$ variables and $rn$ clauses.
For every truth assignment $\sigma\in \{0, 1\}^{n}$ and every clause $c=\ell_{1}\vee\cdots\vee\ell_{k}$,
let $d=d(\sigma, c)$ be the number of satisfied literal occurrences in $c$ under $\sigma$.
For fixed $\beta>-1$ and $\lambda>0$, we take
\begin{align}
\nonumber
\omega(\sigma, c)\propto\left\{
                    \begin{array}{cl}
                      0 & d=0, \\
                      \lambda(1+\beta) & d=1, \\
                      \lambda^{d} & \text{otherwise}.\\
                    \end{array}
                  \right.
\end{align}
Applying the above weighting scheme, we get that if $F_{k}(n, rn)$ is unsatisfiable with probability tending to one as $n\rightarrow\infty$,
then $r\geq2.83$, 8.09, 18.91, 40.81, 84.87 for $k=3$, 4, 5, 6 and 7, respectively.

\end{abstract}

\maketitle

\section{Introduction}

Let $V$ be a set of $n$ boolean variables. A $k$-clause is a disjunction of $k$ boolean variables or their negations.
Let $C_{k}(V)$ denote the set of all $2^{k}n^{k}$ possible $k$-clauses on $V$.
A $k$-CNF formula $F_{k}(n, rn)$ is formed by selecting uniformly, independently and with replacement $rn$ clauses from $C_{k}(V)$ and taking their conjunction \cite{R1, R3, R4}.

If $k$ is allowed to grow with $n$, Frieze and Wormald \cite{R3} proved that if $k-\log_{2}n\rightarrow +\infty$, then random $k$-SAT has a sharp threshold around $n(2^{k}+O(1))\ln 2$.
A few years later, the authors of this paper \cite{R4} relaxed the condition $k-\log_{2}n\rightarrow +\infty$ to $k\geq \left(\frac{1}{2}+\epsilon\right)\log_{2} n$ for any fixed $\epsilon>0$.

For each fixed $k\geq 2$ (i.e., independent of $n$), let
\begin{align}
\nonumber
r_{k}=\sup\left\{r: \lim_{n\rightarrow\infty}\text{\bf Pr}\big[F_{k}(n, rn)\ \text{is satisfiable}\big]=1\right\}.
\end{align}

For every truth assignment $\sigma\in \{0, 1\}^{n}$ and every clause $c=\ell_{1}\vee\cdots\vee\ell_{k}$,
let $d=d(\sigma, c)$ be the number of satisfied literal occurrences in $c$ under $\sigma$.
Fix $\lambda>0$ and let
\begin{align}
\omega(\sigma, c)\propto\left\{
                    \begin{array}{cl}
                      0 & d=0, \\
                      \lambda^{d} & \text{otherwise}.\\
                    \end{array}
                  \right.\label{v1}
\end{align}

Applying the above weighting scheme, Achlioptas and Peres \cite{R1} proved the following result.
{
\flushleft {\bf Theorem 1.1}.\ $There$ $exists$ $a$ $sequence$ $\delta_{k}\rightarrow 0$ $such$ $that$ $for$ $all$ $k\geq 3$,
\begin{align}
\nonumber
r_{k}\geq 2^{k}\ln 2-(k+1)\frac{\ln 2}{2}-1-\delta_{k}.
\end{align}
}

This is the first rigorous proof of a replica method prediction for any NP-complete problem at zero temperature,
and for $k\geq 4$ this improves all previously known lower bounds for $r_{k}$.

In this paper, we propose a novel weighting scheme, which is a revised version of (\ref{v1})
\begin{align}
\omega(\sigma, c)\propto\left\{
                    \begin{array}{cc}
                      0 & d=0, \\
                      \lambda(1+\beta) & d=1, \\
                      \lambda^{d} & \text{otherwise},\\
                    \end{array}
                  \right.\label{v2}
\end{align}
where $\beta>-1$ and $\lambda>0$ are fixed.

By choosing $\beta$ and $\lambda$ properly, we will prove that $r_{3}\geq 2.83$, $r_{4}\geq 8.09$, $r_{5}\geq 18.91$, $r_{6}\geq 40.81$ and $r_{7}\geq 84.87$
so sharpening the lower bounds $r_{3}\geq 2.68$, $r_{4}\geq 7.91$, $r_{7}\geq 84.82$ obtained in \cite{R1}, and $r_{5}\geq 18.79$, $r_{6}\geq 40.74$ obtained by using the same method as in \cite{R1}.

\section{The Second Moment Method}

For a non-negative random variable $X$, making use of the second moment $\text{E}[X^{2}]$ is called the $second$ $moment$ $method$.
In this paper, we use the second moment method in the following form.

{
\flushleft {\bf Lemma 2.1}.\ $For$ $any$ $non$-$negative$ $random$ $variable$ $X$,
\begin{align}
\text{\bf Pr}[X>0]\geq\frac{\text{\bf E}[X]^{2}}{\text{\bf E}[X^{2}]}.\label{2}
\end{align}
}

An attractive feature of the second moment method is that for any non-negative random variable $Y$, if $Y>0$ implies that $X>0$, then
\begin{align}
\text{\bf Pr}[X>0]\geq\text{\bf Pr}[Y>0]\geq\frac{\text{\bf E}[Y]^{2}}{\text{\bf E}[Y^{2}]}.\label{2}
\end{align}

In a breakthrough paper, Friedgut \cite{R5} proved the existence of a $non$-$uniform$ threshold for random $k$-SAT.
{
\flushleft {\bf Theorem 2.2}.\ $For$ $each$ $k\geq 2$, $there$ $exists$ $a$ $sequence$ $r_{k}(n)$ $such$ $that$ $for$ $every$ $\epsilon>0$,
\begin{align}
\nonumber
\lim_{n\rightarrow\infty}\text{\bf Pr}\left[F_{k}(n, rn)\ \text{is satisfiable}\right]=\left\{
                                                        \begin{array}{cc}
                                                          1 & r=(1-\epsilon)r_{k}(n), \\
                                                          0 & r=(1+\epsilon)r_{k}(n). \\
                                                        \end{array}
                                                      \right.
\end{align}
}

Given a $k$-CNF formula F on $n$ variables let $S(F)=\{\sigma: \sigma\ \text{satisfies}\ F\}\subseteq \{0, 1\}^{n}$ denote the set of satisfying truth
assignments of $F$ and let $X=X(F)\geq 0$ be such a random variable that $X>0$ implies that $S(F)\neq\emptyset$.
Sums of the form
\begin{align}
\nonumber
X=\sum_{\sigma}\omega(\sigma, F)
\end{align}
clearly has this property if $\omega(\sigma, F)\geq 0$ and $\omega(\sigma, F)>0$ implies that $\sigma\in S(F)$.

An immediate corollary of Theorem 2.2 is as follows.
{
\flushleft {\bf Corollary 2.3}.\ $For$ $each$ $k\geq 2$, $if$ $\liminf_{n\rightarrow\infty}\text{\bf Pr}\big[F_{k}(n, rn)\ \text{is satisfiable}\big]>0$, $then$ $r_{k}\geq r$.
}

Thus, if for any fixed $r>0$ we have $\text{\bf E}[X^{2}]=O(\text{\bf E}[X]^{2})$, then $r_{k}\geq r$.

Since $F$ is formed by some independent clauses, it is natural to require that $\omega(\sigma, F)$ has product structure over these clauses
\begin{align}
\nonumber
\omega(\sigma, F)=\prod_{c}\omega(\sigma, c),
\end{align}
then clause-independent allows one to replace expectations of products with products of expectations
\begin{align}
\nonumber
\text{\bf E}[\omega(\sigma, F)]=\prod_{c}\text{\bf E}[\omega(\sigma, c)].
\end{align}
With this in mind, let us consider random variables of the form
\begin{align}
X=\sum_{\sigma}\prod_{c}\omega(\sigma, c),\label{b1}
\end{align}
where $\omega(\sigma, c)\geq 0$ and $\omega(\sigma, c)=0$ if $\sigma$ falsifies $c$.

For every truth assignment $\sigma$ and every clause $c=\ell_{1}\vee\cdots\vee\ell_{k}$, we require that $\omega(\sigma, c)=\omega(\text{\bf v})$,
where $\text{\bf v}=(v_{1},\cdots,v_{k})$, $v_{i}=+1$ if $\ell_{i}$ is satisfied under $\sigma$ and $-1$ if $\ell_{i}$ is falsified under $\sigma$.
Since every $\ell_{i}$ in $c$ has the equal right, it is natural to require that $\omega(\text{\bf v})=\omega(|\text{\bf v}|)$, where $|\text{\bf v}|$ denote the number of $+1$s in $\text{\bf v}$.

Let $A=\{-1, +1\}^{k}$ and let $\alpha=z/n$. Then \cite{R1}
\begin{align}
\text{\bf E}[X]=2^{n}\left(\sum_{\text{\bf v}\in A}\omega(\text{\bf v})2^{-k}\right)^{rn},\
\text{\bf E}[X^{2}]=2^{n}\sum_{z=0}^{n}{n \choose z}f_{\omega}(\alpha)^{rn},
\end{align}
where $f_{\omega}(\alpha)=\sum_{\text{{\bf u},{\bf v}}\in A}\omega(\text{\bf u}) \omega(\text{\bf v})2^{-k}\prod_{i=1}^{k}\left(\alpha^{{\bf 1}_{u_{i}=v_{i}}}(1-\alpha)^{{\bf 1}_{u_{i}\neq v_{i}}}\right)$.

The proof of the following Lemma follows by applying the Laplace method of asymptotic analysis \cite{R6}.
{
\flushleft {\bf Lemma 2.4}.\ $Let$ $\phi$ $be$ $any$ $positive$ $function$ $on$ $[0, 1]$ $and$ $let$
\begin{align}
\nonumber
S_{n}=\sum_{z=0}^{n}{n \choose z}\phi(\alpha)^{n}.
\end{align}

$Letting$ $0^{0}=1$, $define$ $g$ $on$ $[0, 1]$ $as$
\begin{align}
\nonumber
g(\alpha)=\frac{\phi(\alpha)}{\alpha^{\alpha}(1-\alpha)^{1-\alpha}}.
\end{align}
$If$ $there$ $exists$ $\alpha_{\max}\in (0, 1)$ $such$ $that$ $g(\alpha_{\max})\equiv g_{\max}>g(\alpha)$ $for$ $all$ $\alpha$ $\neq$ $\alpha_{\max}$ $and$
\begin{align}
\nonumber
g^{''}(\alpha_{\max})= -\frac{g_{\max}}{\alpha_{\max}(1-\alpha_{\max})}\times\rho^{-2} \Big(\phi\ is\ twice\ differentiable\ at\ \alpha_{\max}\Big),
\end{align}
$where$ $\rho>0$, $then$

\begin{align}
\nonumber
\lim_{n\rightarrow\infty}S_{n}/g_{\max}^{n}=\rho.
\end{align}
}

With Lemma 2.4 in mind, let us define
\begin{align}
\nonumber
\Lambda_{\omega}(\alpha)=\frac{2f_{\omega}(\alpha)^{r}}{\alpha^{\alpha}(1-\alpha)^{1-\alpha}}.
\end{align}
Observe that
\begin{align}
\Lambda_{\omega}(1/2)^{n}=\left(4f_{\omega}(1/2)^{r}\right)^{n}=\text{\bf E}[X]^{2}.\label{c1}
\end{align}
Then, Lemma 2.4 and (\ref{c1}) can be combined to yield that if $\Lambda_{\omega}$ has a unique global maximum at $1/2$ on $[0, 1]$ and $\Lambda_{\omega}^{''}(1/2)<0$,
then $\text{\bf E}[X^{2}]=O\left(\text{\bf E}[X]^{2}\right)$.
And we know that \cite{R1}
\begin{align}
\Lambda_{\omega}^{'}(1/2)=0\Longleftrightarrow \sum_{\text{\bf v}\in A}\omega(\text{\bf v})(2|\text{\bf v}|-k)=0.\label{c2}
\end{align}

{\bf The Specific Calculations of Weighting Scheme (\ref{v2})}.
For our weighting scheme, as defined in (\ref{v2}), we can rewrite the right side of (\ref{c2}) as
\begin{align}
\nonumber
  &\sum_{j=1}^{k}{k\choose j}\lambda^{j}(2j-k)+k(2-k)\lambda\beta\\
\nonumber
= &k\left(1-(k-2)\lambda\beta-(1+\lambda)^{k-1}(1-\lambda)\right)\\
\nonumber
= &0,
\end{align}
i.e.,
\begin{align}
(1+\lambda)^{k-1}(1-\lambda)+(k-2)\lambda\beta=1.
\end{align}

For every truth assignment $\sigma$ and every clause $c=\ell_{1}\vee\cdots\vee\ell_{k}$ ($\ell_{1}, \cdots, \ell_{k}$ are i.i.d.) let $S_{1}(c)=\{\sigma: d(\sigma, c)=1\}$
and let $H(\sigma, c)$ be the number of satisfied literal occurrences in $c$ under $\sigma$ less the number of unsatisfied literal occurrences in $c$ under $\sigma$.
For any $\gamma>0$, let
\begin{align}
X=\sum_{\sigma}\prod_{c}\gamma^{H(\sigma, c)}\left({\bf 1}_{\sigma\in S(c)}+\beta\times {\bf 1}_{\sigma\in S_{1}(c)}\right).\label{e}
\end{align}
(Note that $\gamma^{H(\sigma, c)}=\gamma^{2d(\sigma, c)-k}$, so this is consistent with (\ref{v2}) for $\gamma^{2}=\lambda$.)

Let $\sigma$, $\tau$ be any pair of truth assignments that agree on $z=\alpha n$ variables. Then
\begin{align}
\nonumber
   &{\text{\bf E}}\left[\gamma^{H(\sigma, c)+H(\tau, c)}\right]=\left(\alpha\left(\frac{\gamma^{2}+\gamma^{-2}}{2}\right)+1-\alpha\right)^{k},\\
\nonumber
   &{\text{\bf E}}\left[\gamma^{H(\sigma, c)+H(\tau, c)}{\bf 1}_{\sigma\not\in S(c)}\right]={\bf \text{E}}\left[\gamma^{H(\sigma, c)+H(\tau, c)}{\bf 1}_{\tau\not\in S(c)}\right]=\left(\frac{\alpha\gamma^{-2}+1-\alpha}{2}\right)^{k},\\
\nonumber
   &{\text{\bf E}}\left[\gamma^{H(\sigma, c)+H(\tau, c)}{\bf 1}_{\sigma, \tau\not\in S(c)}\right]=\left(\frac{\alpha\gamma^{-2}}{2}\right)^{k},\\
\nonumber
   &{\text{\bf E}}\left[\gamma^{H(\sigma, c)+H(\tau, c)}{\bf 1}_{\sigma\in S_{1}(c)}\right]={\bf \text{E}}\left[\gamma^{H(\sigma, c)+H(\tau, c)}{\bf 1}_{\tau\in S_{1}(c)}\right]\\
\nonumber
=  &{k \choose 1}\left(\frac{\alpha\gamma^{2}+1-\alpha}{2}\right)\left(\frac{\alpha\gamma^{-2}+1-\alpha}{2}\right)^{k-1},\\
\nonumber
   &{\text{\bf E}}\left[\gamma^{H(\sigma, c)+H(\tau, c)}{\bf 1}_{\sigma\not\in S(c), \tau\in S_{1}(c)}\right]={\bf \text{E}}\left[\gamma^{H(\sigma, c)+H(\tau, c)}{\bf 1}_{\sigma\in S_{1}(c), \tau\not\in S(c)}\right]\\
\nonumber
=  &{k \choose 1}\left(\frac{1-\alpha}{2}\right)\left(\frac{\alpha\gamma^{-2}}{2}\right)^{k-1},\\
\nonumber
   &{\text{\bf E}}\left[\gamma^{H(\sigma, c)+H(\tau, c)}{\bf 1}_{\sigma, \tau\in S_{1}(c)}\right]\\
\nonumber
=  &{k \choose 1}\left(\frac{\alpha\gamma^{2}}{2}\right)\left(\frac{\alpha\gamma^{-2}}{2}\right)^{k-1}+2!{k\choose 2}\left(\frac{1-\alpha}{2}\right)^{2}\left(\frac{\alpha\gamma^{-2}}{2}\right)^{k-2}.
\end{align}
Observe that
\begin{align}
\nonumber
  &\left({\bf 1}_{\sigma\in S(c)}+\beta\times {\bf 1}_{\sigma\in S_{1}(c)}\right)\left({\bf 1}_{\tau\in S(c)}+\beta\times {\bf 1}_{\tau\in S_{1}(c)}\right)\\
\nonumber
= &\left({\bf 1}-{\bf 1}_{\sigma\not\in S(c)}+\beta\times {\bf 1}_{\sigma\in S_{1}(c)}\right)\left({\bf 1}-{\bf 1}_{\tau\not\in S(c)}+\beta\times {\bf 1}_{\tau\in S_{1}(c)}\right)\\
\nonumber
= &{\bf 1}-{\bf 1}_{\sigma\not\in S(c)}-{\bf 1}_{\tau\not\in S(c)}+{\bf 1}_{\sigma, \tau\not\in S(c)}+\\
\nonumber
  &\beta\times\left({\bf 1}_{\sigma\in S_{1}(c)}+{\bf 1}_{\tau\in S_{1}(c)}-{\bf 1}_{\sigma\in S_{1}(c), \tau\not\in S(c)}-{\bf 1}_{\sigma\not\in S(c), \tau\in S_{1}(c)}\right)+\\
  &\beta^{2}\times {\bf 1}_{\sigma, \tau\in S_{1}(c)}\equiv\Gamma\big((\sigma, \tau), \beta, c\big).\label{d1}
\end{align}
Therefore
\begin{align}
\nonumber
       &\text{\bf E}\left[\gamma^{H(\sigma, c)+H(\tau, c)}\Gamma((\sigma, \tau), \beta, c)\right]\\
\nonumber
=      &A(\alpha, \gamma)+2k\times B(\alpha, \gamma)\times\beta+k\times C(\alpha, \gamma)\times \beta^{2}\\
\equiv &f(\alpha, \beta, \gamma),\label{d2}
\end{align}
where
\begin{align}
\nonumber
A(\alpha, \gamma)= &\left(\alpha\left(\frac{\gamma^{2}+\gamma^{-2}}{2}\right)+1-\alpha\right)^{k}-2\left(\frac{\alpha\gamma^{-2}+1-\alpha}{2}\right)^{k}+\left(\frac{\alpha\gamma^{-2}}{2}\right)^{k},\\
\nonumber
B(\alpha, \gamma)= &\left(\frac{\alpha\gamma^{2}+1-\alpha}{2}\right)\left(\frac{\alpha\gamma^{-2}+1-\alpha}{2}\right)^{k-1}-\left(\frac{1-\alpha}{2}\right)\left(\frac{\alpha\gamma^{-2}}{2}\right)^{k-1},\\
\nonumber
C(\alpha, \gamma)= &\left(\frac{\alpha\gamma^{2}}{2}\right)\left(\frac{\alpha\gamma^{-2}}{2}\right)^{k-1}+(k-1)\left(\frac{1-\alpha}{2}\right)^{2}\left(\frac{\alpha\gamma^{-2}}{2}\right)^{k-2}.
\end{align}
Then, (\ref{d1}) and (\ref{d2}) can be combined to yield that
\begin{align}
\nonumber
\text{\bf E}[X^{2}]=&\text{\bf E}\left[\sum_{\sigma, \tau}\prod_{c}\gamma^{H(\sigma, c)+H(\tau, c)}\Gamma\big((\sigma, \tau), \beta, c\big)\right]\\
\nonumber
=&\sum_{\sigma, \tau}\prod_{c}\text{\bf E}\left[\gamma^{H(\sigma, c)+H(\tau, c)}\Gamma\big((\sigma, \tau), \beta, c\big)\right]\\
=&2^{n}\sum_{z=0}^{n}{n \choose z}f(\alpha, \beta, \gamma)^{rn}.\label{e2}
\end{align}

With Lemma 2.4 in mind, let us define
\begin{align}
\nonumber
G_{r}(\alpha, \beta, \gamma)=\frac{f(\alpha, \beta, \gamma)^{r}}{\alpha^{\alpha}(1-\alpha)^{1-\alpha}}.
\end{align}
Fix $\beta$ and $\gamma$. Then, if $G_{r}$ has a unique global maximum at $\alpha=1/2$ on $[0, 1]$ and
\begin{align}
\nonumber
\frac{\partial^{2} G_{r}}{\partial \alpha^{2}}\big(1/2, \beta, \gamma\big)<0,
\end{align}
then we get $r_{k}\geq r$.

{\bf An Enhanced Method: Truncation and Weighting.}
For a given $k$-CNF formula $F$ on $n$ variables, recall that $S=S(F)\subseteq \{0, 1\}^{n}$ is the set of satisfying truth assignment of $F$.
Let $S^{+}=\{\sigma\in S: H(\sigma, F)\geq 0\}$ and let $X_{+}=\sum_{\sigma\in S^{+}}\prod_{c}\omega(\sigma, c)$.
For any weighting scheme of $k$-SAT: $\omega(\sigma, c)\geq 0$ and $\omega(\sigma, c)=0$ if $\sigma$ falsifies $c$, we have \cite{R7}
{
\flushleft {\bf Lemma 2.5}. $If$ $f_{\omega}^{'}(1/2)=0$, $then$ $\lim_{n\rightarrow\infty}\text{\bf E}[X_{+}]/\text{\bf E}[X]=1/2$.
}

For any $\gamma>0$, let
\begin{align}
X_{+}=\sum_{\sigma\in S^{+}}\prod_{c}\gamma^{H(\sigma, c)}\left({\bf 1}_{\sigma\in S(c)}+\beta\times {\bf 1}_{\sigma\in S_{1}(c)}\right).
\end{align}
(This is  consistent with those random variables $X$ defined in (\ref{e}), i.e., consistent with our weighting scheme (\ref{v2}).)

Observe that
\begin{align}
&f_{\omega}^{'}(1/2)=0\Longleftrightarrow(1+\gamma^{2})^{k-1}(1-\gamma^{2})+(k-2)\gamma^{2}\beta=1.\label{eqn 1}
\end{align}

A simple calculation gives
\begin{align}
\nonumber
X_{+}^{2}=&\left(\sum_{\sigma\in S^{+}}\prod_{c}\omega(\sigma, c)\right)\left(\sum_{\tau\in S^{+}}\prod_{c}\omega(\tau, c)\right)\\
\nonumber
=&\left(\sum_{\sigma}{\bf 1}_{\sigma\in S^{+}}\prod_{c}\omega(\sigma, c)\right)\left(\sum_{\tau}{\bf 1}_{\tau\in S^{+}}\prod_{c}\omega(\tau, c)\right)\\
\nonumber
=&\sum_{\sigma, \tau}{\bf 1}_{\sigma, \tau\in S^{+}}\prod_{c}\omega(\sigma, c)\omega(\tau, c)\\
=&\sum_{\sigma, \tau}{\bf 1}_{\sigma, \tau\in S^{+}}\prod_{c}\gamma^{H(\sigma, c)+H(\tau, c)}\Gamma\big((\sigma, \tau), \beta, c\big).\label{g1}
\end{align}

Given a tuple $(\beta_{0}, \gamma_{0})\in (-1, +\infty)\times(0, +\infty)$ satisfies the right side of (\ref{eqn 1}).
In particular, if
\begin{align}
\nonumber
X_{+}=\sum_{\sigma\in S^{+}}\prod_{c}\gamma_{0}^{H(\sigma, c)}\left({\bf 1}_{\sigma\in S(c)}+\beta_{0}\times {\bf 1}_{\sigma\in S_{1}(c)}\right),
\end{align}
then for any $\gamma\geq\gamma_{0}$, following the derivation of (\ref{e2}) and (\ref{g1}), we deduce that
\begin{align}
\nonumber
\text{\bf E}[X_{+}^{2}]=&\sum_{\sigma, \tau}\text{\bf E}\left[{\bf 1}_{_{\sigma, \tau\in S^{+}}}\prod_{c}\gamma_{0}^{H(\sigma, c)+H(\tau, c)}\Gamma\big((\sigma, \tau), \beta_{0}, c\big)\right]\\
\nonumber
\leq&\sum_{\sigma, \tau}\text{\bf E}\left[\prod_{c}\gamma^{H(\sigma, c)+H(\tau, c)}\Gamma\big((\sigma, \tau), \beta_{0}, c\big)\right]\\
\nonumber
=&\sum_{\sigma, \tau}\text{\bf E}\left[\gamma^{H(\sigma, c)+H(\tau, c)}\Gamma\big((\sigma, \tau), \beta_{0}, c\big)\right]^{rn}\\
=&2^{n}\sum_{z=0}^{n}{n \choose z}f(\alpha, \beta_{0}, \gamma)^{rn}.
\end{align}
Therefore
\begin{align}
\text{\bf E}[X_{+}^{2}]\leq 2^{n}\sum_{z=0}^{n}{n \choose z}\inf_{\gamma\geq\gamma_{0}}f(\alpha, \beta_{0}, \gamma)^{rn}\equiv 2^{n}\sum_{z=0}^{n}{n \choose z}\underline{f}(\alpha, \beta_{0}, \gamma_{0})^{rn}.
\end{align}

With Lemma 2.4 in mind and we take
\begin{align}
g_{r}(\alpha, \beta_{0}, \gamma_{0})=\frac{\underline{f}(\alpha, \beta_{0}, \gamma_{0})^{r}}{\alpha^{\alpha}(1-\alpha)^{1-\alpha}}\left(\leq\frac{f(\alpha, \beta_{0}, \gamma_{0})^{r}}{\alpha^{\alpha}(1-\alpha)^{1-\alpha}}= G_{r}(\alpha, \beta_{0}, \gamma_{0})\right).
\end{align}

Note now that if $g_{r}(1/2, \beta_{0}, \gamma_{0})>g_{r}(\alpha, \beta_{0}, \gamma_{0})$ for all $\alpha\neq 1/2$,
then
\begin{align}
\nonumber
g_{r}(1/2, \beta_{0}, \gamma_{0})=G_{r}(1/2, \beta_{0}, \gamma_{0}),
\end{align}
otherwise,
let $\varepsilon=G_{r}(1/2, \beta_{0}, \gamma_{0})-g_{r}(1/2, \beta_{0}, \gamma_{0}) (>0)$,
then there exists a constant $D>0$ such that for all sufficiently large $n$,
\begin{align}
\nonumber
\text{\bf E}[X_{+}^{2}]\leq   &\big(2g_{r}(1/2, \beta_{0}, \gamma_{0})+\varepsilon\big)^{n}\leq\big(2g_{r}(1/2, \beta_{0}, \gamma_{0})+2\varepsilon\big)^{n}\\
\nonumber
=                             &\big(2G_{r}(1/2, \beta_{0}, \gamma_{0})\big)^{n}=\text{\bf E}[X]^{2}\leq D\times\text{\bf E}[X_{+}]^{2}\\
\nonumber
\leq                          &D\times\text{\bf E}[X_{+}^{2}].
\end{align}
Note that
\begin{align}
\nonumber
\left(\frac{2g_{r}(1/2, \beta_{0}, \gamma_{0})+2\varepsilon}{2g_{r}(1/2, \beta_{0}, \gamma_{0})+\varepsilon}\right)^{n}\rightarrow+\infty\ \text{as}\ n\rightarrow\infty,
\end{align}
and this is a contradiction.

Suppose that $g_{r}(1/2, \beta_{0}, \gamma_{0})>g_{r}(\alpha, \beta_{0}, \gamma_{0})$ for all $\alpha\neq 1/2$ and
\begin{align}
\nonumber
\frac{\partial^{2} G_{r}}{\partial \alpha^{2}}\big(1/2, \beta_{0}, \gamma_{0}\big)<0.
\end{align}
Note that there exists a constant $\epsilon$ such that $G_{r}(1/2, \beta_{0}, \gamma_{0})>G_{r}(\alpha, \beta_{0}, \gamma_{0})$ hold on for all $\alpha\in (1/2-\epsilon, 1/2)\cup(1/2, 1/2+\epsilon)$.
With Lemma 2.4 in mind, we consider the function defined as follows:
\begin{align}
\nonumber
\phi(\alpha, \beta_{0}, \gamma_{0})=\left\{
                   \begin{array}{ll}
                     f(\alpha, \beta_{0}, \gamma_{0}) &\text{if}\ \alpha\in(1/2-\epsilon, 1/2+\epsilon),\\
                     \underline{f}(\alpha, \beta_{0}, \gamma_{0}) &\text{otherwise}.\\
                   \end{array}
                 \right.
\end{align}
It is clear that
\nonumber
\begin{align}
\text{\bf E}[X_{+}^{2}]\leq\sum_{z=0}^{n}{n \choose z}\phi(\alpha, \beta_{0}, \gamma_{0})^{rn},
\end{align}
then by Lemma 2.4, we get $r_{k}\geq r$.

\section{Use of the Method}
We apply the method to the case $k=$3, 4, 5, 6 and 7.
To do this, we demonstrate values $\beta$, $r$ and let $\gamma$ satisfies the right side of (\ref{eqn 1}) such that $g_{r}(1/2, \beta, \gamma)>g_{r}(\alpha, \beta, \gamma)$ hold on for all $\alpha\neq 1/2$ and
\begin{align}
\nonumber
\frac{\partial^{2} G_{r}}{\partial \alpha^{2}}\big(1/2, \beta, \gamma\big)<0.
\end{align}

We obtain
\begin{center}
\begin{tabular}{cccccc}
$k$         &3        &4     &5     &6     &7\\
\hline
$\beta$     &$(0.56, 0.74)$     &$(0.13, 0.15)$  &$(0.04, 0.06)$  &0.02  &0.01\\
$r$         &2.83     &8.09  &18.91 &40.81 &84.87\\
\end{tabular}
\end{center}

\end{document}